\documentclass[twocolumn,amsmath,superscriptaddress,amssymb,aps,pra]{revtex4-1}

\usepackage{graphicx,txfonts,enumerate}
\usepackage[colorlinks, linkcolor=blue, citecolor=blue, urlcolor=blue, breaklinks=true]{hyperref}

\newcommand{\eref}[1]{Eq.~(\ref{#1})}
\newcommand{\erefs}[1]{Eqs.~(\ref{#1})}
\newcommand{\fref}[1]{Fig.~\ref{#1}}
\newcommand{\frefs}[1]{Figs.~\ref{#1}}
\newcommand{\sref}[1]{Sec.~\ref{#1}}

\DeclareMathOperator{\Sym}{Sym}

\begin{document}

\title{Thermodynamics of trajectories and local fluctuation theorems for harmonic quantum networks}

\author{Simon Pigeon}
\email[Corresponding author:\ ]{s.pigeon@qub.ac.uk}
\affiliation{Centre for Theoretical Atomic, Molecular and Optical Physics, School of Mathematics and Physics, Queen's University Belfast, Belfast BT7\,1NN, United Kingdom}
\author{Lorenzo Fusco}
\affiliation{Centre for Theoretical Atomic, Molecular and Optical Physics, School of Mathematics and Physics, Queen's University Belfast, Belfast BT7\,1NN, United Kingdom}
\author{Andr\'e Xuereb}
\affiliation{Department of Physics, University of Malta, Msida MSD2080, Malta}
\affiliation{Centre for Theoretical Atomic, Molecular and Optical Physics, School of Mathematics and Physics, Queen's University Belfast, Belfast BT7\,1NN, United Kingdom}
\author{Gabriele De Chiara}
\affiliation{Centre for Theoretical Atomic, Molecular and Optical Physics, School of Mathematics and Physics, Queen's University Belfast, Belfast BT7\,1NN, United Kingdom}
\author{Mauro Paternostro}
\affiliation{Centre for Theoretical Atomic, Molecular and Optical Physics, School of Mathematics and Physics, Queen's University Belfast, Belfast BT7\,1NN, United Kingdom}

\date{\today}

\begin{abstract}
We present a general method to undertake a thorough analysis of the thermodynamics of the quantum jump trajectories followed by an arbitrary quantum harmonic network undergoing linear and bilinear dynamics. The approach is based on the phase-space representation of the state of a harmonic network. The large deviation function associated with this system encodes the full counting statistics of exchange and also allows one to deduce for fluctuation theorems obeyed by the dynamics. We illustrate the method showing the validity of a local fluctuation theorem about the exchange of excitations between a restricted part of the environment (i.e., a local bath) and a harmonic network coupled with different schemes.
\end{abstract}


\maketitle

\section{Introduction}
The recent development of thermodynamics of trajectories for quantum systems promises to shed new light on the thermodynamics of quantum systems \cite{Garrahan2010,Budini2011,Hickey2012,Pigeon2015}. Based on the density matrix representation of a system it allows, through the large deviation function, to access the full counting statistics of the exchange of excitations between a system and its environment but also to explore the long-time behaviour of a system, revealing phenomena such as dynamical phase transitions \cite{Hickey2012,Pigeon2014}. However, similarly to thermodynamics of trajectories for classical systems \cite{Touchette2009}, the effectiveness of the method is usually limited by the practical difficulty of obtaining the large deviation function. As the density matrix of the system reaches a long-time limit, the method of thermodynamics of trajectories requires, in principle, significant computational effort to find the large deviation function. Furthermore for a system evolving in an infinite-dimensional Louiville space the situation is even more difficult because necessary truncation of the space will be needed, leading to an approximated large deviation function.\\
In this article we will present a general method for the determination of the large deviation function for a large variety of systems evolving in an infinite-dimensional Louiville space, allowing a sensible reduction of computational power and without need of approximations. We present a method for the full characterisation of the exchange of excitations with the environment for a general network of quantum harmonic oscillators, undergoing linear and bilinear dynamics. Our method is based on two essential steps: (i)~A quantum-optic phase-space representation of the network's degrees of freedom, and (ii)~a multidimensional Gaussian ansatz.

We will present the general framework in
\sref{sec:2}, introducing the large deviation function that encodes
 the statistics for the exchange of excitations. The method which we will detail can handle
all possible linear and bilinear interactions. In \sref{sec:3}
we will show how this method can be used to numerically verify local detailed
fluctuation theorems on the exchange with a given bath. We will first consider the simplest case of
a single harmonic oscillator, whose large deviation function has recently been analytically derived~\cite{Pigeon2015}
using a similar approach (\sref{sub:32-1}) followed in \sref{sub:32} by a harmonic chain where the inter-oscillator coupling is rotating-wave-like (RW). Following this, we will consider two coupled oscillators where each is damped by a given thermal bath with a squeezing-like (\sref{sub:34}) and position--position-like (\sref{sub:35}) inter-oscillator coupling.

\section{General framework\label{sec:2}}

In this section we will detail the derivation of the large deviation
function for an arbitrary network of quantum harmonic oscillators.
We will start by defining our model (\sref{sub:21}) and
its dynamics, followed by the unraveling of the considered exchange process
and its related thermodynamics (\sref{sub:22}).
In \sref{sub:23} we will present the phase space representation
of the network, and the quantum Fokker--Planck equation derived from it.
Using a Gaussian ansatz we will formally define the large deviation
function (\sref{sub:24}).

\subsection{Modeling a harmonic network\label{sub:21}}

Considering a set of $N$ quantum harmonic oscillators, the
network Hamiltonian can be written as $\hat{H}=\sum_{i=1}^{N}\hat{H}_{i}+\sum_{i>j}^{N}\hat{H}_{ij}$,
in terms of single-oscillator ($\hat{H}_{i}$) and two-oscillator ($\hat{H}_{ij}$) Hamiltonians. For simplicity, and because of our later restriction to Hamiltonians that are at most quadratic in the operators, we restrict ourselves to bipartite coupling between oscillators. As we will be interested only in linear and bilinear Hamiltonians, we can explicitly write the single-oscillator Hamiltonians as
\begin{multline}
\hat{H}_{i}=\hbar\omega_{i}\left(\hat{a}_{i}^{\dagger}\hat{a}_{i}+\tfrac{1}{2}\right)+\hbar d_{i}(t)\omega_{i}\left(\hat{a}_{i}^{\dagger}+\hat{a}_{i}\right)+(\hbar\varUpsilon_{i}\hat{a}_{i}\hat{a}_{i}+h.c.).
\label{eq:H}
\end{multline}
This corresponds to the Hamiltonian of a harmonic oscillator, of frequency
$\omega_{i}$, driven by a bounded time-dependent force $\vert d_{i}(t)\vert\le D_{i}$,
undergoing single mode squeezing with rate $\varUpsilon_{i}\in\mathbb{C}$. The coupling between oscillators encoded through $\hat{H}_{ij}$ can take three different forms
\begin{enumerate}[(i)]
\item The position--position coupling ($x$--$x$ type)
\begin{equation}
\hat{H}_{ij}^{xx}=\hbar g_{ij}\left(\hat{a}_{i}+\hat{a}_{i}^{\dagger}\right)\left(\hat{a}_{j}+\hat{a}_{j}^{\dagger}\right) \,,\label{eq:XX}
\end{equation}

\item The rotating-wave (RW) coupling
\begin{equation}
\hat{H}_{ij}^{\text{RW}}=\hbar g_{ij}\left(\hat{a}_{i}\hat{a}_{j}^{\dagger}+\hat{a}_{i}^{\dagger}\hat{a}_{j}\right)\label{eq:RWA}\,,
\end{equation}

\item The two-mode squeezing (OPO-like) coupling
\begin{equation}
\hat{H}_{ij}^{\text{OPO}}=\hbar g_{ij}\left(\hat{a}_{i}^{\dagger}\hat{a}_{j}^{\dagger}+\hat{a}_{i}\hat{a}_{j}\right)\label{eq:OPO}\,,
\end{equation}
\end{enumerate}
where in each case we have $g_{ij}=g_{ji}$. The dynamics of the system is given by the master equation $\partial_{t}\hat{\rho}=\mathcal{W}[\hat{\rho}]$,
where $\hat{\rho}$ is the density matrix of the full network and the superoperator
$\mathcal{W}[\bullet]=-\imath\bigl[\hat{H},\bullet\bigr]+\mathcal{L}[\bullet]$ describes the system dynamics. $\mathcal{L}=\sum_{i=1}^{N}\left(\mathcal{L}_{i}+\mathcal{S}_{i}\right)$ is
the global dissipator composed of two types of exchange channels: number damping channels and squeezing damping channels, respectively described by the superoperators
\begin{multline}
\mathcal{L}_{i}[\bullet]=\bar{\Gamma}_{i}\left(2\hat{a}_{i}^{\dagger}\bullet\hat{a}_{i}-\left\{ \bullet,\hat{a}_{i}\hat{a}_{i}^{\dagger}\right\} \right)+\Gamma_{i}\left(2\hat{a}_{i}\bullet\hat{a}_{i}^{\dagger}-\left\{ \bullet,\hat{a}_{i}^{\dagger}\hat{a}_{i}\right\} \right)\label{eq:lcav-1}\,,
\end{multline}
and
\begin{multline}
\mathcal{S}_{i}[\bullet]=\bar{\Lambda}_{i}\left(2\hat{a}_{i}\bullet\hat{a}_{i}-\left\{ \bullet,\hat{a}_{i}\hat{a}_{i}\right\} \right)
+\Lambda_{i}\left(2\hat{a}_{i}^{\dagger}\bullet\hat{a}_{i}^{\dagger}-\left\{ \bullet,\hat{a}_{i}^{\dagger}\hat{a}_{i}^{\dagger}\right\} \right)\label{eq:lcav-2-1}\,.
\end{multline}
Based on this Lindblad form of the master equation, we will now present our method to unravel the statistics of exchange of excitations between the network and a given bath.

\subsection{Unraveled statistics and thermodynamics of trajectories\label{sub:22}}

To build the trajectories we will follow the approach introduced
in Ref.~\cite{Garrahan2010}. To do so we have to define the
observable of interest for the exchange of excitations between the system and its environment. We introduce a counting process described by the number $K_r$, which gives the number of the quanta exchanged between the system and part of the environment, a given bath $r$, defined as
\begin{equation}
K_{r}:=K_{r-}-K_{r+}\label{eq:k}\,,
\end{equation}
where $K_{r\pm}$ are the numbers of quanta entering and leaving
the oscillator coupled to the bath $r$. We note here that the index $r$ will be used throughout to denote the `reference' bath for which we are studying the exchange statistics.

The probability of obtaining a
given value of $K_r$ after a time $t$ will be defined as $p_{K_r}(t)=\text{Tr}\left\{ \hat{P}^{K_r}\hat{\rho}\right\} $ where $\hat{P}^{K_r}$ is a projector over the subspace associated to $K_r$ excitations. From the probability  $p_{K_r}(t)$ we can define the moment generating function, also known as the dynamical partition function, as
\begin{equation}
Z_r(s,t)=\sum_{K_r} e^{-sK_r}p_{K_r}(t)\,.\label{eq:z}
\end{equation}
From the large deviation theory we know that in the long-time limit we
have $Z(s,t)\sim e^{t\theta_r(s)}$, where $\theta_r(s)$ is the large
deviation function. The large deviation function is the fundamental
building block of the theory of thermodynamics of trajectories and encodes the long-time
dynamics of the system relatively to a given counting process $K_{r}$.\\
Once defined, the counting process number $K_r$ is used to bias the trajectories
as in \eref{eq:z} \cite{Garrahan2010}. A biased density matrix can be defined as $\hat{\rho}_{s}:=\sum_{K_r}e^{-sK_r}\hat{P}^{K_r}\hat{\rho}$, and the corresponding dynamics is given by the biased master equation $\partial_{t}\hat{\rho}_{s}=\mathcal{W}[\hat{\rho}_{s}]+\mathcal{L}_{s}[\hat{\rho}_{s}]$,
where $\mathcal{W}$ is the superoperator associated to the unbiased
system while $\mathcal{L}_{s}$ is the non-trace-preserving part of
the dynamics emerging from the biasing procedure and encoding the statistics
of interest. For the considered counting process we have
\begin{equation}
\mathcal{L}_{s}[\bullet]=2\Gamma_{r}(e^{-s}-1)\hat{a}_{r}\bullet\hat{a}_{r}^{\dagger}+2\bar{\Gamma}_{r}(e^{s}-1)\hat{a}_{r}^{\dagger}\bullet\hat{a}_{r}\,.\label{eq:ls-2}
\end{equation}\\
The large deviation function $\theta_r(s)$ can be defined with respect to the biased density matrix $\hat{\rho}_{s}$ as:
\begin{equation}
\theta_r(s)=\lim_{t\to\infty}\frac{1}{t}\ln\left[\text{Tr}\left\{ \hat{\rho}_{s}\right\} \right]\label{eq:ll}
\end{equation}
where the index $r$ refers to reference bath. In order to solve the above equation we will now consider the phase-space representation of the system.

\subsection{Phase-space representation and the generating function\label{sub:23}}

The phase-space representation is a well-established method commonly used
in quantum mechanics to deal with quantum harmonic oscillators \cite{Gardiner2010,Carmichael2002}.
The advantage of this approach is that a harmonic oscillator, evolving
along an infinite Hilbert space, can be fully characterised by means of
a quasi-probability distribution evolving in the complex plane. In
what follows we will concentrate on the characteristic function associated
with this quasi-probability. We will consider the symmetrically-ordered
generating function
\begin{equation}
\chi_{s}(\beta_{1},\dots,\beta_{N})=\text{Tr}\left\{ \exp\left[i\sum_{i=1}^{N}\left(\beta_{i}^{*}\hat{a}_{i}^{\dagger}+\beta_{i}\hat{a}_{i}\right)\right]\hat{\rho}_{s}\right\}\,,\label{eq:generate}
\end{equation}
but a similar approach can be conducted with other
representations.
Details of the derivation of the phase-space representation of different parts of the dynamics can be found in Appendix. We can collect the different contributions to the dynamics in term of the complex coordinates $\beta_{i}=p_{i}+iq_{i}$, writing the quantum Fokker--Planck equation, associated with the generating function $\chi_{s}$, in the following form
\begin{multline}
\partial_{t}\chi_{s}=\Bigl[\mathbf{p}^{T}\cdot\mathbf{A}\cdot\partial_{\mathbf{p}}+\mathbf{p}^{T}\cdot\mathbf{D}\cdot\mathbf{p}+\mathbf{d}^{T}\cdot\mathbf{p}\\
+\tfrac{1}{2}\left(\partial_{\mathbf{p}}^{T}\cdot\mathbf{F}_{s}^{+}\cdot\partial_{\mathbf{p}}+\mathbf{p}^{T}\cdot\mathbf{F}_{s}^{+}\cdot\mathbf{p}\right)\\
-\left(\mathbf{p}^{T}\cdot\mathbf{F}_{s}^{-}\cdot\partial_{\mathbf{p}}+\tfrac{1}{2}\text{Tr}\left\{ \mathbf{F}_{s}^{-}\right\} \right)\Bigr]\,\chi_{s}\label{eq:xs}\,.
\end{multline}
 Here $\mathbf{p}^{T}=\left(p_{1},q_{1},...,p_{N},q_{N}\right)$ is
the vector $p_{i}$ and $q_{i}$ conjugate fields of respectively the position and momentum quadratures. The first line of \eref{eq:xs} refers to the
unbiased part of the dynamics, given by the superoperator $\mathcal{W}$,
while the second and third refer to the biased part, given by $\mathcal{L}_{s}$.
The drift matrix $\mathbf{A}$ is defined
as
\begin{equation}
\mathbf{A}=\bigoplus_{i=1}^{N}\begin{pmatrix}-2\Im[\Upsilon_{i}]-\Gamma_{i}+\bar{\Gamma}_{i} & -\omega_{i}+2\Re[\Upsilon_{i}]\\
\omega_{i}+2\Re[\Upsilon_{i}] & 2\Im[\Upsilon_{i}]-\Gamma_{i}+\bar{\Gamma}_{i}
\end{pmatrix}+\mathbf{G}\,,
\end{equation}
where $\mathbf{G}$ is the coupling matrix, which  can be written as
\begin{equation}
\mathbf{G}=\begin{pmatrix}0 & \mathbf{G}_{2,1} & \cdots & \mathbf{G}_{N,1}\\
\mathbf{G}_{1,2} & 0 & \cdots & \mathbf{G}_{N,2}\\
\vdots & \vdots & \ddots & \vdots\\
\mathbf{G}_{1,N} & \mathbf{G}_{2,N} & \cdots & 0
\end{pmatrix}
\end{equation}
with $\mathbf{G}_{i,j}=\mathbf{G}_{j,i}$ and the following  coupling scheme-dependent definitions
\begin{equation}
\mathbf{G}_{i,j}=\begin{cases}
\begin{pmatrix}0 & 0\\
2g_{ij} & 0
\end{pmatrix}_{xx}&\text{($x$--$x$ type),}\\
\begin{pmatrix}0 & -g_{ij}\\
g_{ij} & 0
\end{pmatrix}_{\text{RW}}&\text{(RW type),}\\
\begin{pmatrix}0 & g_{ij}\\
g_{ij} & 0
\end{pmatrix}_{\text{OPO}}&\text{(OPO-like type).}
\end{cases}\label{eq:gij}
\end{equation}
In Eq.~\eqref{eq:xs}, $\mathbf{D}$ is the noise matrix defined as
\begin{equation}
\mathbf{D}=\bigoplus_{i=1}^{N}\begin{pmatrix}-\Gamma_{i}-\bar{\Gamma}_{i}+2\Re(\Lambda_{i}) & -2\Im(\Lambda_{i})\\
-2\Im(\Lambda_{i}) & -\Gamma_{i}-\bar{\Gamma}_{i}-2\Re(\Lambda_{i})
\end{pmatrix}\,.
\end{equation}
Finally, for what concerns the unbiased part of the dynamics we have
the driving vector
$\mathbf{d}^T=\left(  0,\omega_{1}d_{1}(t),\cdots,0,\omega_{N}d_{N}(t) \right)^T$.
With these definitions we
can describe all the processes addressed so far.

The second and third lines of \eref{eq:xs} account for the biased part of the dynamics. Indeed, we have
\begin{equation}
\mathbf{F}_{s}^{\pm}=\bigoplus_{ji=1}^{N}\delta_{ir}\begin{pmatrix}f_{i\pm}(s) & 0\\
0 & f_{i\pm}(s)
\end{pmatrix},
\end{equation}
where $r$ labels the reference bath, and
$f_{i\pm}(s)=\Gamma_{i}(e^{-s}-1)\pm\bar{\Gamma}_{i}(e^{s}-1)$.
We can now rewrite ~\eref{eq:ll} in terms of the generating function $\chi_{s}$ as
\begin{equation}
\theta_r(s)=\lim_{t\to\infty}\frac{1}{t}\ln\left[\chi_{s}(\mathbf{0})\right]\,.\label{eq:ll-2}
\end{equation}
This is possible owing to $\text{Tr}\left\{ \hat{\rho}_{s}\right\} =\chi_{s}(\mathbf{0})$,
where $\chi_{s}(\mathbf{0})$ is the volume of the
biased quasi-probability distribution, i.e., the biased Wigner function
in the present case. We remark that this quantity is not dependent on the choice of the specific type of phase-space
representation. Notice that the above definition is valid for any harmonic network, subjected to an arbitrary dynamical process. We now restrict our attention to linear and bilinear processes in order to proceed further with our analysis in a fully analytical form.

\subsection{Gaussian ansatz and large deviation function\label{sub:24}}

To solve \eref{eq:ll-2} we now consider a multidimensional Gaussian ansatz. Its validity relies on the fact that, when undergoing linear dynamics, a Gaussian state remains Gaussian at all times. This argument can be easily extended to non-Gaussian initial conditions converging with time to Gaussian states, as described by the central limit theorem.
This allows us to formulate the problem in terms of the finite number of parameters entering the Gaussian ansatz. Considering multiple coupled harmonic oscillators, each associated to a two-dimensional phase space (generated by $p_{i}$ and $q_{i}$), our ansatz reads
\begin{equation}
\chi_{s}=A_{s}\exp\left(\imath\,\mathbf{p}^{T}\cdot\mathbf{x}_{s}-\tfrac{1}{2}\mathbf{p}^{T}\cdot\mathbf{\Sigma}_{s}\cdot\mathbf{p}\right)\,,\label{eq:ansatz1-1}
\end{equation}
where $\mathbf{x}_{s}^{T}=\left(x_{1},y_{1},...,x_{N},y_{N}\right)$
is the vector of expectation values of position and momentum of each oscillator (here $k_i=\langle\hat{k}_{i}\rangle$ with $k=x,y$). The covariance matrix $\mathbf{\Sigma}_{s}$ can be decomposed in terms of the two-dimensional block matrices as
\begin{equation}
\mathbf{\Sigma}_{s}=\begin{pmatrix}\mathbf{\Sigma}_{1,1} & \mathbf{\Sigma}_{1,2} & \cdots & \mathbf{\Sigma}_{1,N}\\
\mathbf{\Sigma}_{2,1} & \mathbf{\Sigma}_{2,2} & \cdots & \mathbf{\Sigma}_{2,N}\\
\vdots & \vdots & \ddots & \vdots\\
\mathbf{\Sigma}_{N,1} & \mathbf{\Sigma}_{N,2} & \cdots & \mathbf{\Sigma}_{N,N}
\end{pmatrix}\,,
\end{equation}
where
$\mathbf{\Sigma}_{i,j}=\begin{pmatrix}\sigma_{i,j}^{xx} & \sigma_{i,j}^{xy}\\
\sigma_{i,j}^{yx} & \sigma_{i,j}^{yy}
\end{pmatrix}$,
and $\sigma_{i,j}^{ef}=\tfrac{1}{2}\langle\bigl(\hat{e}_{i}\hat{f}_{j}+\hat{f}_{j}\hat{e}_{i}\bigr)\rangle-\langle\hat{e}_{i}\rangle\langle\hat{f}_{j}\rangle$ ($\hat{e}_{i},\hat{f}_{i}\in\{\hat{x}_{i},\hat{y}_{i}\}$). By definition, we have $\mathbf{\Sigma}_{i,j}=\mathbf{\Sigma}_{j,i}^{T}$.
As the biased density matrix $\hat{\rho}_{s}$ is not normalised, we have
to take into account the norm of the generating function $A_{s}$. Moreover $A_s$ is the central quantity of interest since we notice, from
\eref{eq:ll-2}, that the large deviation function is given by
\begin{equation}
\theta_r(s)=\lim_{t\to\infty}\frac{1}{t}\ln\left[A_{s}(t)\right]\label{eq:ll-3}\,.
\end{equation}

Applying this ansatz to \eref{eq:xs} we can extract the
time evolution of the norm $A_{s}$, the position/momentum vector
$\mathbf{x}_{s}$, and the covariance matrix $\mathbf{\Sigma}_{s}$.
Notice that the $s$ index illustrates the dependence of these elements
upon the bias parameter $s$. We find that
\begin{equation}
2\partial_t\ln A_s(t)
=\text{Tr}\left\{ \mathbf{F}_{s}^{+}\cdot\mathbf{\Sigma}_{s}(t)\right\} +\mathbf{x}_{s}(t)^{T}\cdot\mathbf{F}_{s}^{+}\cdot\mathbf{x}_{s}(t)-\text{Tr}\left\{ \mathbf{F}_{s}^{-}\right\} \label{eq:a-1}\,.
\end{equation}
For the first moment we have
\begin{equation}
\mathbf{\dot{x}}_{s}(t)=\left[\mathbf{A}-\mathbf{F}_{s}^{-}+\mathbf{F}_{s}^{+}\cdot\mathbf{\Sigma}_{s}(t)\right]\cdot\mathbf{x}_{s}(t)+\mathbf{d}(t)\label{eq:xx-2}\,,
\end{equation}
and for the second
\begin{align}
\mathbf{\dot{\Sigma}}_{s}(t)&=\left(\mathbf{A}-\mathbf{F}_{s}^{-}\right)\cdot\mathbf{\Sigma}_{s}(t)+\mathbf{\Sigma}_{s}(t)\cdot\left(\mathbf{A}-\mathbf{F}_{s}^{-}\right)^{T}\nonumber\\
&+\mathbf{\Sigma}_{s}(t)\cdot\mathbf{F}_{s}^{+}\cdot\mathbf{\Sigma}_{s}(t)+\mathbf{F}_{s}^{+}-2\mathbf{D}\label{eq:sigma-1}\,.
\end{align}\\
\erefs{eq:a-1}-(\ref{eq:sigma-1}) define the evolution of the generating function at any time, governed by the biased
master equation. To obtain the unbiased dynamics we simply have to
take $s\to0$, or equivalently $\mathbf{F}_{s}^{\pm}\to\mathbf{0}$.
 Going one step further, using \eref{eq:a-1}, we have that the large deviation function is
\begin{multline}
\theta_r(s)=\lim_{t\to\infty}\frac{1}{2t}\int_{0}^{t}\bigl[\text{Tr}\left\{ \mathbf{F}_{s}^{+}{\cdot}\mathbf{\Sigma}_{s}(\tau){-}\mathbf{F}_{s}^{-}\right\}+\mathbf{x}_{s}^{T}(\tau)\cdot\mathbf{F}_{s}^{+}\cdot\mathbf{x}_{s}(\tau)\bigr]d\tau\,.\label{eq:mvtheta-1-3}
\end{multline}
This definition is valid for any harmonic network undergoing linear and bilinear processes. From the counting process
considered here and the associated matrix $\mathbf{F}_{s}^{\pm}$, we find
\begin{multline}
\theta_r(s)=\lim_{t\to\infty}\frac{1}{2t}\int_{0}^{t}\bigl\{f_{r+}(s)\sum_{k=x,y}\bigl[\sigma_{r}^{k}(s,t)+k_{r}^{2}(s,t)\bigr]-2f_{r-}(s)\bigr\}\,d\tau\,,\label{eq:mvtheta-1-3-1}
\end{multline}
where the means and variances are here dependent on time $t$, and on
the bias parameters $s$, while $r$ refers to the bath under consideration.

It is interesting now to look at some specific cases. Let us assume that the
system converges towards a stationary state, in which case $\lim_{t\to\infty}\mathbf{\Sigma}_{s}(t)=\tilde{\mathbf{\Sigma}}_{s}$ with
$\tilde{\mathbf{\Sigma}}_{s}$ the covariance matrix of the stationary solution of \eref{eq:sigma-1}. We find that
\begin{multline}
\theta_r(s)=\frac{1}{2}\text{Tr}\left\{ \mathbf{F}_{s}^{+}{\cdot}\mathbf{\tilde{\Sigma}}_{s}{-}\mathbf{F}_{s}^{-}\right\}+\lim_{t\to\infty}\frac{1}{2t}\int_{0}^{t}\left[\mathbf{x}_{s}^{T}(\tau){\cdot}\mathbf{F}_{s}^{+}{\cdot}\mathbf{x}_{s}(\tau)\right]d\tau\label{eq:mvtheta-1-2-1}\,.
\end{multline}

As we are ultimately interested in the long-time behaviour,
a simple approximation can be used to obtain
the last term of \eref{eq:mvtheta-1-2-1} without the need of the full time evolution
of $\mathbf{x}_{s}(t)$ and $\mathbf{\Sigma}_{s}(t)$. It consists
in replacing the stationary covariance matrix $\tilde{\mathbf{\Sigma}}_{s}$ in the evolution equation of $\mathbf{x}_{s}(t)$ [\eref{eq:xx-2}].
This approach was used in Ref.~\cite{Pigeon2015}
to solve analytically the large deviation of a driven harmonic
oscillator coupled to $N$ baths.

In the more restrictive case where the Hamiltonian is quadratic in creation and annihilation operators (i.e., we have no driving),  \eref{eq:xx-2}
achieves a stationary solution with $\mathbf{x}_{s}=\mathbf{0}$. Therefore, the last
term in \eref{eq:mvtheta-1-3} drops, leaving a large deviation
function depending only on the stationary biased covariance matrix
$\mathbf{\tilde{\Sigma}}_{s}$ as
\begin{equation}
\theta_r(s)=\frac{1}{2}\text{Tr}\left\{ \mathbf{F}_{s}^{+}\cdot\mathbf{\tilde{\Sigma}}_{s}-\mathbf{F}_{s}^{-}\right\}.\label{eq:mvtheta-1-2-1-1}
\end{equation}

We have thus seen how, through \erefs{eq:xx-2}
and (\ref{eq:sigma-1}), we can obtain the
large deviation function associated to a given counting process $K_{r}$, this
being the net number of excitations exchange with the $r^{\rm th}$ bath in contact with the system,
as long as the oscillators undergo linear and bilinear dynamics. Moreover, assuming the existence of stationary
solutions, the complexity of the problem dramatically reduces. In this case, we do not need the full system evolution,
but only its stationary solution given by the large deviation approach. Here the problem
corresponds to solving an algebraic Riccati equation [\eref{eq:sigma-1}]
for different values of the bias parameters $s$. This type of algebraic
equation is frequently encountered in dynamical control problems. Accurate numerical methods exist
to solve this type of equation [cf.  Ref.~\cite{Laub1979} for formal approaches to the solution of a Riccati equation].
Through a phase-space approach complemented by a suitable Gaussian ansatz,
we have presented  a powerful method to access the large deviation
function exactly. As stated above the large deviation function encodes by definition the full counting statistics, since
$\partial_{s}^{n}\theta_r(s)\vert_{s=0}=\left(-1\right)^{n}\kappa_{n}$,
where $\kappa_{n}$ is the $n$-th cumulant of the counting
process $K_r$ \cite{Garrahan2010,Hickey2012,Budini2011,Pigeon2015,Pigeon2014}. However this function encodes other crucial thermodynamic information about the system, one example of which is given by the possibility to formulate fluctuation theorems. Our method gives access to this invaluable source of information, for a large variety of quantum systems, through reasonable computational effort, as we will now demonstrate.

\section{Local fluctuation theorems\label{sec:3}}

We now introduce the
general concept of a fluctuation theorem (FT) (\sref{sub:31})
and its connection with the thermodynamics of trajectories. In Sec.~\ref{sub:32-1} and \ref{sub:32} we focus on a
single harmonic oscillator and a harmonic chain, respectively, and determine the
associated FT. We show how our approach is able to recover the results of previous investigations and to go beyond them by providing an explicit route for the approach of physically relevant forms of coupling among the oscillators belonging to a given network (cf. \sref{sub:34} and \ref{sub:35}).

\subsection{Fluctuation theorems in thermodynamics of trajectories\label{sub:31}}
Fluctuation theorems (FTs) are used to fully characterise the fluctuations endured by a system while interacting with an environment \cite{Lebowitz1999,Evans2002}. Several FTs can be formulated, depending on the scenario considered. We will here focus on FTs for a system in a stationary state. More precisely, we will concentrate on local FTs, related to the exchange between a system and \emph{part} of the environment. The idea is that it is not always possible to keep track of all dissipation processes undergone by the system. In such cases local FTs allow to discuss fluctuation relations in the exchange processes between a system and part of its environment. Moreover, as we will see, while global FTs can be formulated in a wide variety of physical contexts, for local FTs the results are less general. For example, considering the total exchange between a system an its environment, FTs always find a definition \cite{Evans2002,Seifert2005} (eventually through extended version of typical \eref{eq:ft} \cite{Rakos2008}), while in the local case FTs cannot always be found as we will see. This is not dependent on wether the system under consideration is classical or quantum, but it is a consequence of the possibility to have some correlations in the exchange of excitations with different parts of the environment which gives valuable information on the thermodynamical behaviour of the system.

Consider a net number counting process such as $K_{r}:=K_{r-}-K_{r+}$ (where we remind that
$K_{r\pm}$ stands for the net number of quanta leaving and entering into
the system from a specific bath labeled $r$), $p_{K_{r}}(t)$ is the probability to observe a given net number $K_{r}$ of excitations exchanged after a time $t$, and $p_{-K_{r}}(t)$ is the probability
to observe the counting process number $-K_{r}$. If we have
\begin{equation}
\lim_{t\to\infty}\frac{p_{K_{r}}(t)}{p_{-K_{r}}(t)}=e^{K_{r}s_{r}}\label{eq:ft}\,,
\end{equation}
where $s_r$ is independent of $K_r$, then a local fluctuation theorem exists with respect to the $r^{\rm th}$ bath. We know that the existence of a fluctuation theorem is, by definition, associated
with the existence of a specific Gallavotti--Cohen symmetry relation of the large deviation
function $\theta_{r}(s)$ related to the counting process $K_{r}$~\cite{Lebowitz1999}. This can be written explicitly as
\begin{equation}
\theta_{r}(s)=\theta_{r}(s_{r}-s)\label{eq:sym},
\end{equation}
where the symmetric point $s_{r}$ is given in \eref{eq:ft}. The derivation of the latter can be quite involved, whereas determining the existence of symmetry properties of a function can be done efficiently, making the large deviation function a powerful tool to determine FTs.
In order to illustrate the opportunity embodied by the method presented here, in relation to the determination of fundamental thermodynamic relations, we will now focus on the simple example of a single quantum harmonic oscillator.

\subsection{Example 1: Quantum harmonic oscillator\label{sub:32-1}}

For a single harmonic oscillator of frequency $\omega$ coupled to multiple
baths, the large deviation function can be obtained analytically
using the method introduced in Ref.~\cite{Pigeon2015} and highlighted here.
We can access the exchange statistics between the system and a
given bath, considering the counting process $K_{r}$ as previously defined.
In this case we find that the symmetric point of the large deviation function $\theta_r(s)$ is given by
\begin{equation}
s_{r}=\ln\left[\frac{\Gamma_{r}}{\bar{\Gamma}_{r}}\frac{\sum_{i\neq r}^{N}\bar{\Gamma}_{i}}{\sum_{i\neq r}^{N}\Gamma_{i}}\right]\label{eq:sym-1}\,,
\end{equation}
where $\Gamma_{i}$ ($\bar{\Gamma}_{i}$) refers to the rate of exchange of excitations from (to) the system to (from) the $i^{\rm th}$ bath. First of all, in this case we have that a local fluctuation theorem exists in any case and whatever the environment architecture could be. Moreover it depends only on the rate of excitation exchange between the system and the baths, and not on the internal system parameters. To illustrate such features, we consider the simple case of
two thermal baths coupled with the same strength to the system. We have $\Gamma_{i}=(\bar{n}_{i}+1)\gamma/2$ and $\bar{\Gamma}_{i}=\bar{n}_{i}\gamma/2$,
with $\bar{n}_{i}=\left[\exp\left(\hbar\omega/k_\mathrm{B}T_{i}\right)-1\right]^{-1}$
the density of excitations in the $i^{\rm th}$ bath \cite{Gardiner2010}. In this case, the
symmetric point is given by
\begin{equation}
s_{1}=\frac{\hbar\omega}{k_\mathrm{B}}\left(\frac{1}{T_{1}}-\frac{1}{T_{2}}\right),\label{eq:stant}
\end{equation}
which corresponds to the typical entropy flux taking place between two baths at temperatures $T_1$ and $T_2$.

From \eref{eq:sym-1} we can see that with only two connected baths we have $s_{2}=-s_{1}$, indicating that the statistics of exchange
between the system and one bath is strongly related to that
with the second. Considering
that we are addressing a simple scenario where the system cannot store
or transform any of the absorbed excitations, whatever enters the system from one side gets out
from the other side with the same statistics. Consequently we will have an identical and opposite statistics leading to $\theta_1(s)=\theta_2(-s)$, and thus $s_{2}=-s_{1}$. The system
just conducts from one bath to another, with the statistical properties of the exchange of excitations depending on the overall environment. Under these circumstances it becomes clear that the statistics of heat flowing into or out of the system will be the same as the ones revealed by $\theta_1(s)$, while the total net exchange with both baths will be null. As a consequence, we can deduce that the system acts as a perfect thermal conductor. 

As this elementary example shows, determining the large deviation
function can directly lead to the definition of a local fluctuation theorem. The method used here to obtain the large deviation function is similar
to the one developed in the first section, with the nuance that here
exact results can be found because of the simplicity of the system
\cite{Pigeon2015}. For more complex systems a numerical calculation for
the stationary solution of \eref{eq:sigma-1} is necessary.

From now on we will consider different types of harmonic chains where the chain is connected to the environment by its two end oscillators. Each end oscillator is coupled to a thermal bath with identical coupling strength $\gamma$, such that $\Gamma_{i}=(\bar{n}_{i}+1)\gamma/2$ and $\bar{\Gamma}_{i}=\bar{n}_{i}\gamma/2$. One bath will be cold ($T_{1}$) and one hot ($T_{N}>T_{1}$).

\subsection{Example 2: Rotating-wave-type harmonic chain\label{sub:32}}

\begin{figure}
\includegraphics[width=0.35\textwidth]{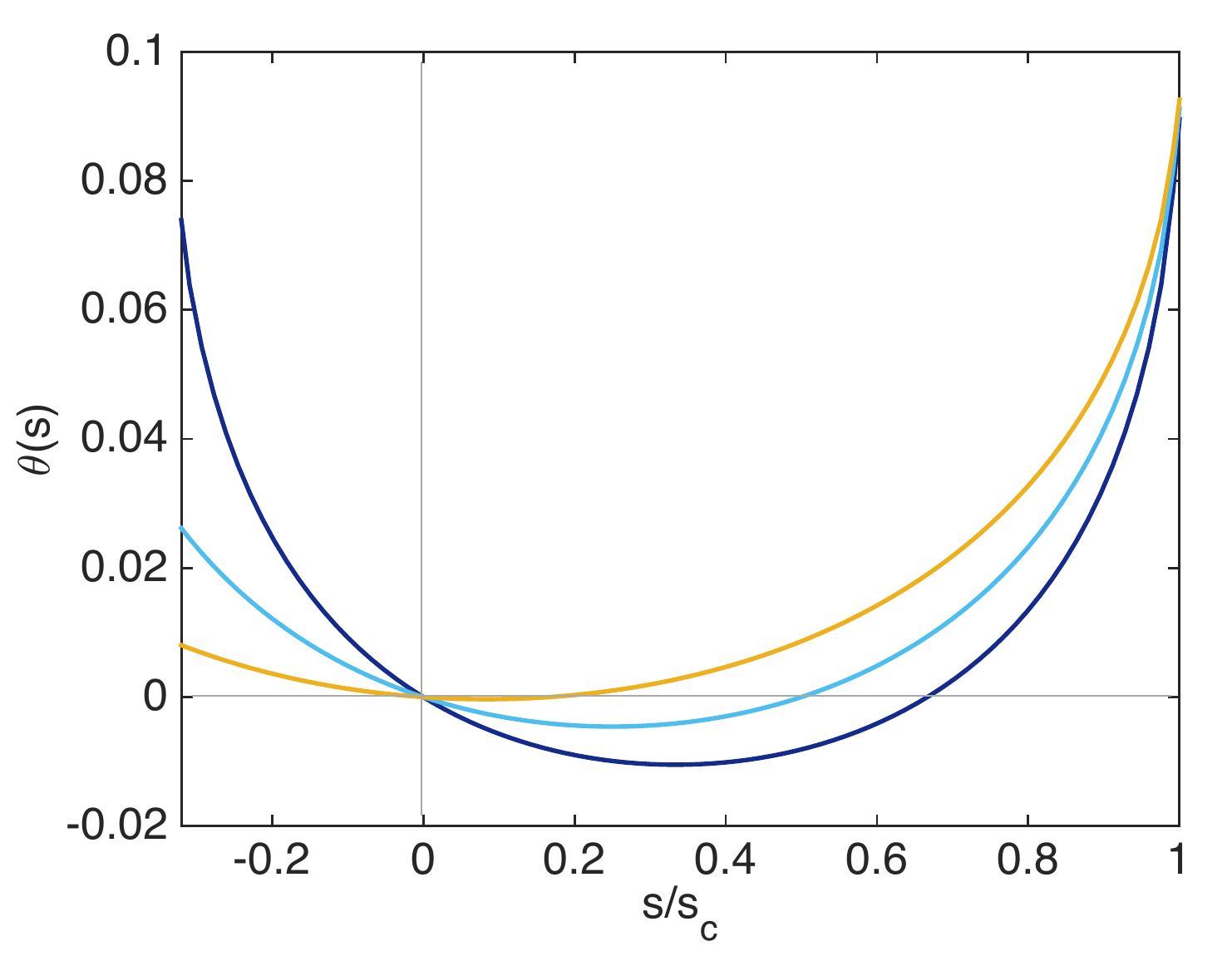}
\caption{Large deviation function $\theta_{1}(s)$ for a harmonic chain of
$N=10$ oscillators. The bias parameter $s$ is normalised with respect to $s_\mathrm{c}=\hbar\omega/k_\mathrm{B}T_{1}$.
Each color corresponds to a different temperature of the (cold) bath $1$: $T_1=0.5\hbar\omega/k_\mathrm{B}$ in dark blue,
$T_1=\hbar\omega/k_\mathrm{B}$ in light blue, and $T_1=5\hbar\omega/k_\mathrm{B}$ in yellow (respectively from lower to
higher for $s>0$). The bath parameters are such that the temperature difference is fixed to $\Delta T=T_{N}-T_{1}=\hbar\omega/k_\mathrm{B}$,
$g=0.1\omega$, and $\gamma=0.1\omega$.\label{fig:theta}}
\end{figure}

Let us consider a chain of $N$ coupled harmonic oscillators of frequency $\omega_i$. Given that only the first and last oscillators are coupled to the environment, the matrices $\mathbf{D}$
and $\mathbf{A}$ are defined as
\begin{equation}
\mathbf{D}=-\frac{\gamma}{2}\bigoplus_{i=1}^{N}\begin{pmatrix}\left(2\bar{n}_{i}+1\right)\left(\delta_{i,1}+\delta_{i,N}\right) & 0\\
0 & \left(2\bar{n}_{i}+1\right)\left(\delta_{i,1}+\delta_{i,N}\right)\label{noise}
\end{pmatrix}
\end{equation}
and
\begin{equation}
\mathbf{A}=\begin{pmatrix}\mathbf{A}_{1} & \mathbf{G}_{1,2} & 0 & \cdots & 0\\
\mathbf{G}_{1,2} & \mathbf{A}_{2} & \mathbf{G}_{2,3} & \cdots & 0\\
0 & \mathbf{G}_{2,3} & \mathbf{A}_{3} & \cdots & 0\\
\vdots & \vdots & \vdots & \ddots & \vdots \\
0 & 0 & 0 & \cdots & \mathbf{A}_{N}
\end{pmatrix}\,,
\end{equation}
with
\begin{equation}
\mathbf{A}_{i}=\begin{pmatrix}-\frac{\gamma}{2}\left(\delta_{i,1}+\delta_{i,N}\right) & -\omega_i\\
\omega_i & -\frac{\gamma}{2}\left(\delta_{i,1}+\delta_{i,N}\right)
\end{pmatrix}\,,
\end{equation}
and with $\mathbf{G}_{i,j}$ the two-oscillator RW-like coupling matrix as defined in
\eref{eq:gij}.

In \fref{fig:theta} we show the large deviation function obtained
from the steady-state solution of \eref{eq:sigma-1} related
to the exchange with the bath $1$ for identical oscillators ($\omega_i=\omega$). The different curves correspond to different bath temperatures $T_{1}$, while $\Delta T=T_N -T_1=\hbar\omega/k_\mathrm{B}$. The bias parameter is normalised with respect
to $s_\mathrm{c}=\hbar\omega/k_\mathrm{B}T_{1}$, where $s=s_\mathrm{c}$ corresponding
to a branch point.

\begin{figure}
\includegraphics[width=0.4\textwidth]{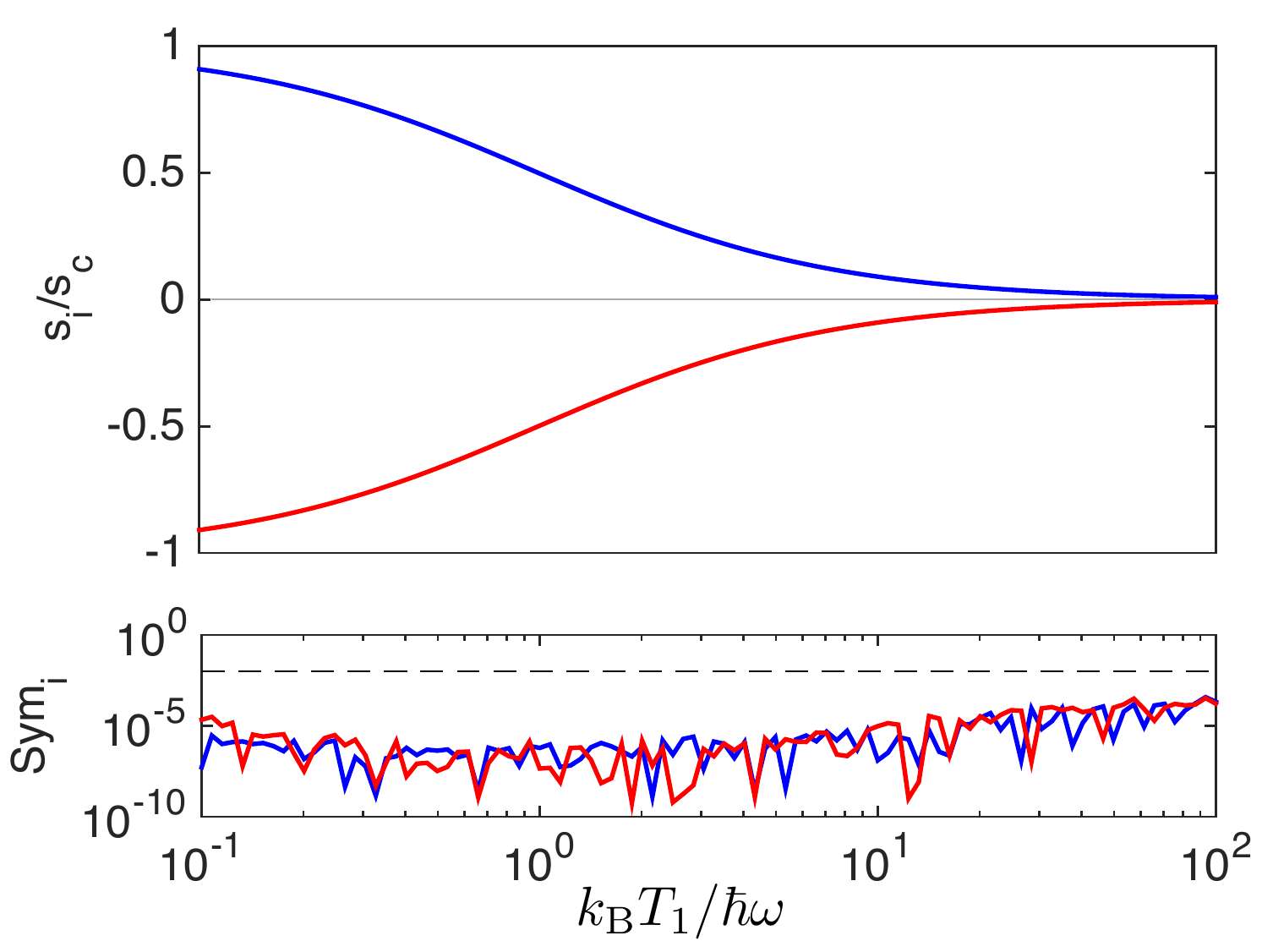}
\caption{Symmetry of the large deviation function $\theta_{i}(s)$. In the
upper panel the symmetric points $s_{i}$ of the large deviation function are shown,
normalised by $s_\mathrm{c}$, and on the lower panel is shown $\Sym_{i}$,
the symmetry criterion of $\theta_i(s)$ as defined in \eref{eq:crit}.
The blue line refers to the exchange with the bath $1$ and the red line
to the exchange with the bath $2$. The dashed line corresponds to
the threshold defined, below which $\theta(s)$ is assumed to be symmetrical.
Both are shown as a function of the bath temperature $T_{1}$. Other parameters
are taken to be the same as for \fref{fig:theta}.\label{fig:sym}}
\end{figure}

Based on the determination of $\theta_1(s)$ as presented in \fref{fig:theta}, the Gallavotti--Cohen symmetry can be obtained, leading to a possible FT.
To that purpose we need to determine $s_{\min}$, the minimum of $\theta_{1}(s)$, and check if it corresponds to an axis of symmetry. Note
that here, imposed by the flexibility of the approach encompassing
a wide variety of systems, the symmetry property of the large deviation function
cannot be directly derived from the definition of $\theta_{r}(s)$ in \eref{eq:mvtheta-1-3}, unlike what was found in Ref.~\cite{Saito2007}
for a related system. In the upper panel of \fref{fig:sym}, we represent the possible local fluctuation theorem
$s_r=2s_{\min}$ between the system and each bath, such as $\theta_{r}(s_{\min})=\min\left(\theta_r(s)\right)$. To determine if $s_{\min}$ is a symmetry point, we define the following quantity
\begin{equation}
\Sym_{r}=\left\vert \frac{\theta_{r}(2s_{\min})}{\theta_{r}(s_{\min})}\right\vert.
\label{eq:crit}
\end{equation}
Should $\theta_r(s)$ be symmetric with respect to $s_{r}=2s_{\min}$ we should have $\theta_{r}(s_{r})=\theta_{r}(0)=0$
by definition, and so $\Sym_{1}\to0$. The behaviour of \eref{eq:crit} against $k_BT_1/\hbar\omega$ is presented in the lower panel of \fref{fig:sym}, where we appreciate that $\Sym_r<10^{-2}$ throughout the whole window of sampled temperatures, thus providing strong numerical evidence of the symmetry of the large deviation function.
A criterion based on $\Sym_{1}\to0$ is a valid test for symmetry whenever $\theta_r(s)$ is continuously differentiable, as in our examples. It allows us to incorporate numerical errors introduced when solving the Riccati equation and provides a qualitative understanding of the behaviour of local FTs.

\fref{fig:sym} shows clearly that (i) a local FT exists at any temperature and that (ii) it behaves exactly as for the single harmonic oscillator case, i.e., $s_{1}=\frac{\hbar\omega}{k_\mathrm{B}}\left(\frac{1}{T_{1}}-\frac{1}{T_{2}}\right)$. The independence
of the FT obtained on the size of the system has to be connected to
the type of coupling between the oscillators, which is responsible for the conservation of the number
of excitations. As soon as
an excitation enters the system another has to exit. This leads
to the same conclusion as before that the system is a perfect heat
conductor because $s_{1}=-s_{2}$, as shown in \fref{fig:sym}.
The convergence to $0$ of $s_1$ and $s_2$ at high temperatures indicate that the system thermalises also locally.
Indeed, for any network of oscillators connected to two baths that is stable in the sense that the eigenvalues of the $\mathbf{A}$ matrix have a all negative imaginary part will give the exact same result, as long as the coupling is of RW-type in between all oscillators.
This independence on the heat conduction on the geometry of the system considered can be the cause for the breakdown of Fourier law observed for harmonic chains \cite{Nicacio}.
We next turn our attention to different kinds of coupling between oscillators.

\subsection{Example 3: Two thermal squeezed modes\label{sub:34}}

\begin{figure}
\includegraphics[width=0.4\textwidth]{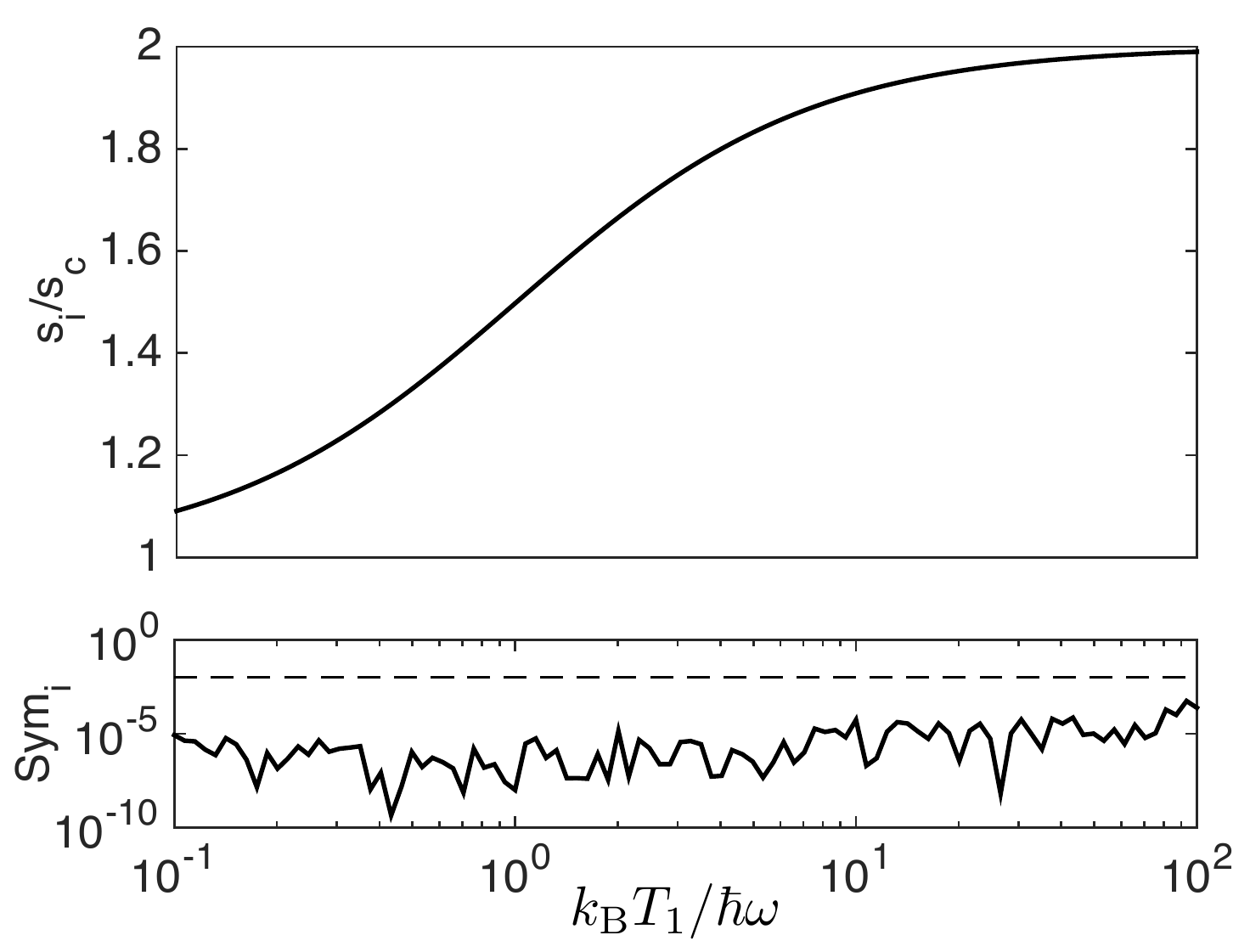}
\caption{Symmetry of the large deviation function $\theta_{1}(s)$ for two oscillators coupled through squeezing. In the upper panel the symmetric point
$s_{1}$ of the large deviation function normalised over $s_\mathrm{c}$
and on the lower panel evaluation of $\Sym_1$, the symmetry criterion in \eref{eq:crit}. The other parameters are as in \fref{fig:theta}, except
the number of oscillators which is reduced to $2$.\label{fig:sym-1}}
\end{figure}

Here we consider an archetypal scenario often encountered in quantum
optics, i.e. two harmonic oscillators interacting through a squeezing Hamiltonian and each connected to a thermal bath.
This model describes a large variety of physical systems from optical parametric amplification \cite{Collett1984} to optomechanical system \cite{Purdy} as in other hybrid quantum system \cite{Mahboob}.
The Hamiltonian is
\begin{equation}
\hat{H}=\hbar\omega\sum_{i=1}^{2}\left(\hat{a}_{i}^{\dagger}\hat{a}_{i}+\frac{1}{2}\right)+\hbar\left(g\hat{a}_{2}\hat{a}_{1}+\text{H.c.}\right)\,,
\end{equation}
where we have simplified our notation by setting $\omega_1=\omega_2=\omega$.
Applying the approach presented previously, we have
\begin{equation}
\mathbf{A}=\begin{pmatrix}-\frac{\gamma}{2} & -\omega & 0 & g\\
\omega & -\frac{\gamma}{2} & g & 0\\
0 & g & -\frac{\gamma}{2} & -\omega\\
g & 0 & \omega & -\frac{\gamma}{2}
\end{pmatrix}
\end{equation}
for the drift matrix, while the noise matrix will be similar to the one previously encountered in \eref{noise}.

Similarly to what has been done previously, we compute the large deviation related to the net
number of excitations exchange between the oscillator $1$ and its
bath. Determining the minimum of this function and evaluating its
symmetry properties we found, as represented in \fref{fig:sym-1},
that (i)~a fluctuation theorem indeed exists and (ii)~it matches the relation $s_{1}=({\hbar\omega}/{k_\mathrm{B}})\left({1}/{T_{1}}+{1}/{T_{2}}\right)$.

Considering now the exchange between the system and the second bath we find that the respective local FTs agreed, such that $s_{2}=s_{1}$. The system operates here emitting heat to both the baths ($s_{r}>0$) with a rate depending on both bath temperatures but independent of inter-oscillator coupling. This independence derive from the system being in a global state (two mode squeezed state) damped through local channels.
This result is in direct contrast with what was observed previously. This behaviour derives from the dissimilarity between the type of inter-oscillator coupling and the one with the baths, leading to a situation where the system cannot thermalise.

Note that the choice of two identical oscillators can and was extended to more oscillators
with the present method. We found that for a chain of identical oscillators
coupled through squeezing coupling with damping on the first and last
oscillators, there exists a stable solution for the system only for an even number of oscillators, while the local FTs remain unchanged.

With these two examples, we have seen two extremely different local FTs and thermodynamic
behaviour. Let us next consider an intermediate example.

\subsection{Example 4: Two oscillators coupled through relative distance \label{sub:35}}

Consider now two oscillators coupled through an  $\left(\hat{x}_{1}-\hat{x}_{2}\right)^{2}$
coupling (such as the motional degrees of freedom of two trapped ions, for example \cite{Cirac1992}). Each
oscillator is in contact with a bath at
a given temperature. The interest in this type of coupling arises from the fact
that it combines both a conservative interaction and a squeezing one,
corresponding to a combination of both the cases presented previously. Due to this combination
the results obtained are quite different to the one obtained before.

We have for the drift matrix $\mathbf{A}$
\begin{equation}
\mathbf{A}=\begin{pmatrix}-\frac{\gamma}{2} & -\omega & 0 & 0\\
\omega+2g & -\frac{\gamma}{2} & -2g & 0\\
0 & 0 & -\frac{\gamma}{2} & -\omega\\
-2g & 0 & \omega+2g & -\frac{\gamma}{2}
\end{pmatrix},
\end{equation}
while $\mathbf{D}$ remains as in \eref{noise}. If previously the outcome of the system and associated FTs were independent
of the system parameters (oscillator frequencies $\omega$ and coupling
strength $g$), here the situation is very different. We will therefore focus on the dependence on two key parameters:~The coupling strength $g$ between oscillators, and~the coupling strength $\gamma$ to the baths.

\subsubsection{Dependence on the inter-oscillator coupling strength $g$}

\begin{figure}
\includegraphics[width=0.4\textwidth]{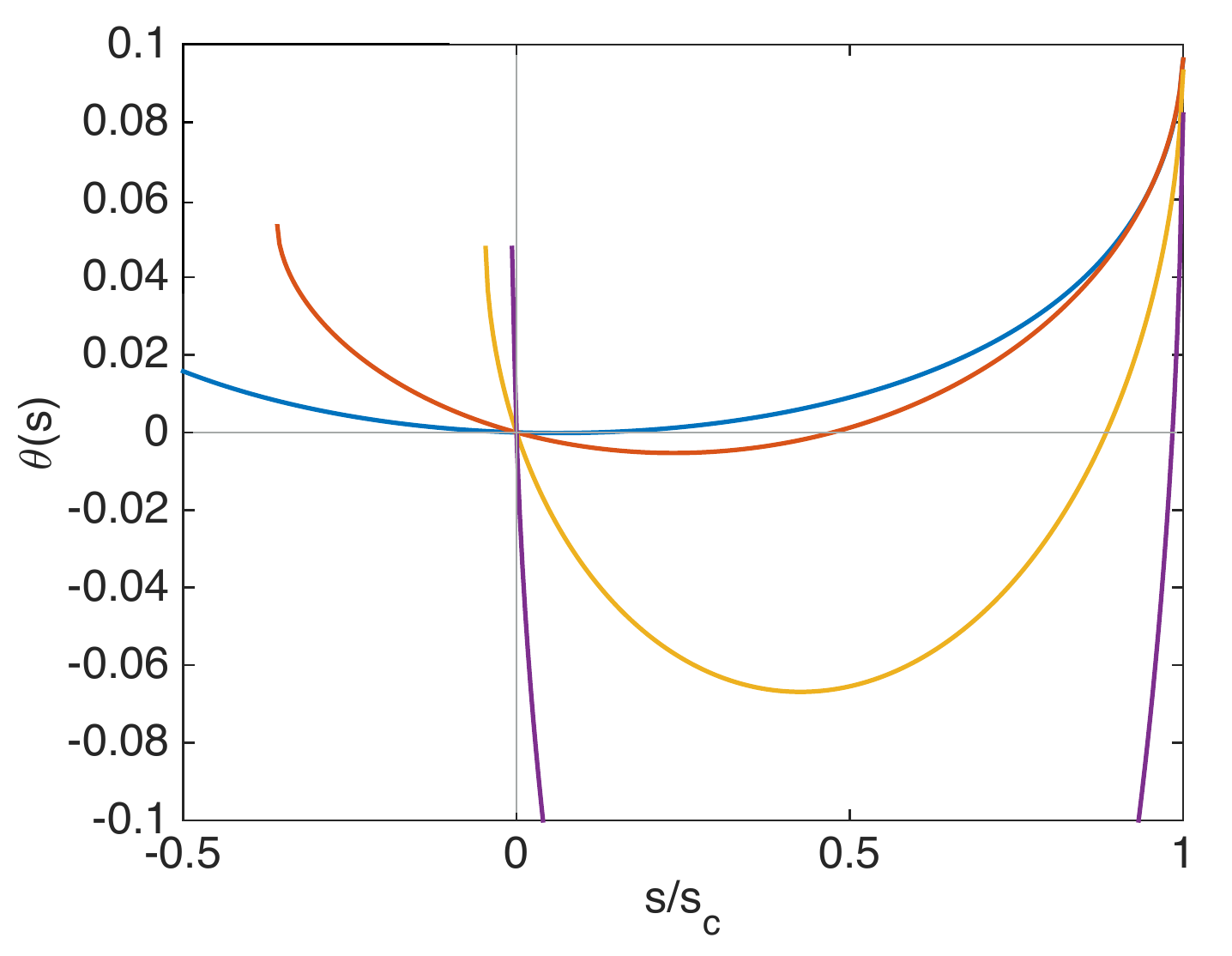}
\caption{Large deviation function for two oscillators coupled through
$\left(\hat{x}_{1}-\hat{x}_{2}\right)^{2}$ coupling for various coupling
strengths $g$. Blue $g=0.1\omega$, orange $g=\omega$, yellow
$g=10\omega$, and purple $g=100\omega$, while $\gamma=0.1\omega$. The bath temperature is $T_{1}=10\hbar\omega/k_\mathrm{B}$, while other parameters are as in \fref{fig:theta}.
\label{fig:thetaxx}}
\end{figure}

\begin{figure}[b]
\includegraphics[width=0.4\textwidth]{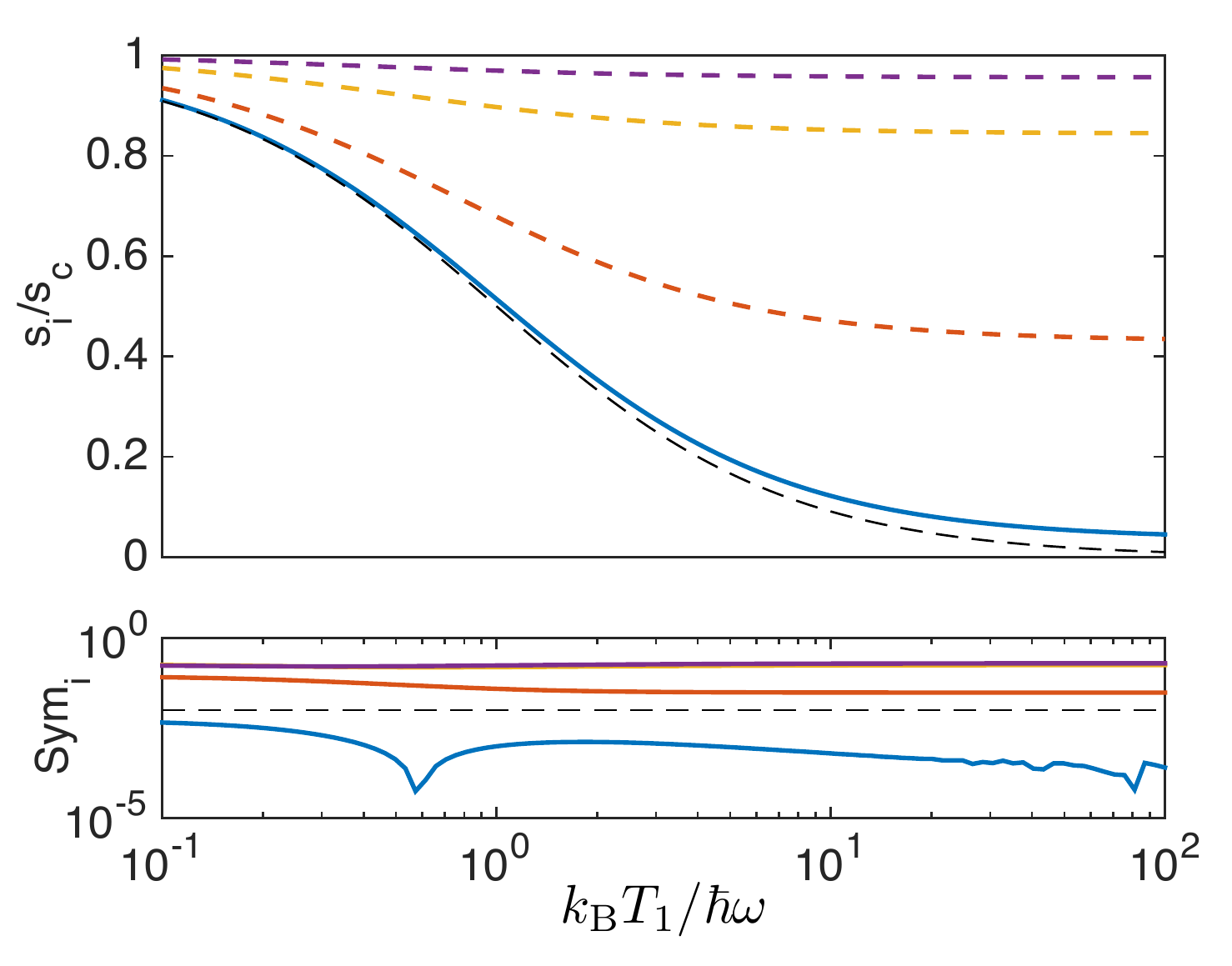}
\caption{Symmetry of the large deviation function $\theta_1(s)$.
In the upper panel the potential FT $s_1=2s_{\min}$ normalised. The thin black
dashed line indicates the result obtained for RW-coupling between oscillators (see section
\ref{sub:32}), while the other curves correspond to various values
of $g$ (same colour code as in \fref{fig:thetaxx}). The full line corresponds
to the situation where $s_{1}$ is found to correspond to a FT, while
the dashed part to situation where is not, relatively to the symmetry criterion $\Sym_1$ [\eref{eq:crit}],
as represented in the lower panel.\label{fig:thetaxx2}}
\end{figure}

In \fref{fig:thetaxx} large deviation functions (related to the exchange with bath $1$) corresponding to various coupling strengths $g$ are shown. For $s>0$ we have
that the smaller the coupling (blue curve), the flatter $\theta_1(s)$. This can be
directly related to changes of the activity or average net number
of quanta exchanged with the bath ($\langle K_r\rangle/t=\partial_{s}\theta_r(s)\vert_{s=0}$), which
increases as the coupling $g$ increases. Looking for FTs, we
can observe in \fref{fig:thetaxx} that increasing the coupling $g$, the minimum $s_{\min}$ seems to pass from $0$ (blue line) to $s_\mathrm{c}=0.5\hbar\omega/k_\mathrm{B}T_{1}$
(purple line). Simultaneously, the branch point located on the negative values
of $s$ tends to $0$.
Finally regarding the possible symmetry properties of $\theta_r(s)$,
the increase of the coupling leads to a non
symmetric function (connected to the change of position of the negative branch point).
On the other hand, when decreasing the coupling, the
negative branch point converges to $-s_\mathrm{c}$, leading to a symmetric large deviation function associated to FT.

We show this behaviour in \fref{fig:thetaxx2}. As previously done
(c.f.\ \frefs{fig:sym} and \ref{fig:sym-1}) the hypothetical FT
$s_{1}=2s_{\min}$ is shown as a function of $T_1$ for different coupling strengths $g$ (same colour code as in \fref{fig:thetaxx}). From these
plots we observe that, by increasing $g$ (from bottom to top
curve), $2s_{\min}$ tends to $s_\mathrm{c}$.
For small coupling (blue, bottom curve) we have that $2s_{\min}$ is close to the case
where the coupling between oscillators was of RW-type (black dashed
line) as expected, presenting however a finite difference. We see that only in the case of small coupling strengths between oscillators we find a local FT.
As previously observed, the existence of a local FT, unlike global FT, is not necessarily guaranteed, as demonstrated here. The non-existence of local FT could be due to partial correlation effects between the statistics of exchange with the different baths.

\begin{figure}
\includegraphics[width=0.4\textwidth]{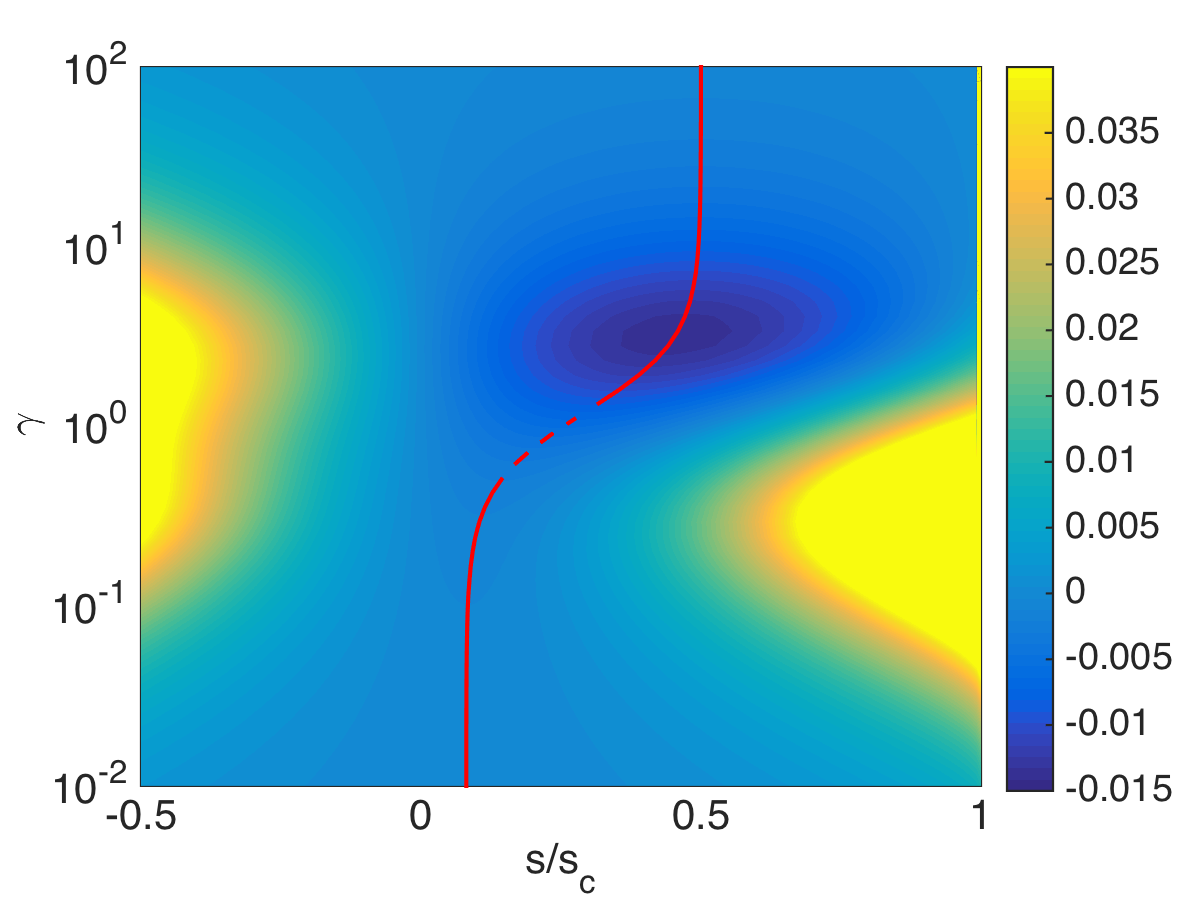}
\caption{Map of the large deviation function $\theta_1(s)$ for various coupling strengths to the baths $\gamma$ (vertical axis) for a temperature of $T_{1}=10\hbar\omega/k_\mathrm{B}$ and $g=0.2\omega$. The red curve
represents $s_{\min}$ and the dashed part indicates when $\theta_1(s)$ is not symmetric with respect to $s_{\min}$ (according to the criterion given in \eref{eq:crit}). Other parameters are the same as in \fref{fig:thetaxx}.\label{fig:thetaxx-1}}
\end{figure}

\subsubsection{Dependence on the coupling strength to the baths $\gamma$}

We now focus on the coupling strength to the baths, $\gamma$,
and how it affects potential FTs. To this end, we
plot in Fig. \ref{fig:thetaxx-1} a map of the
large deviation function versus $s$ and $\gamma$. The red line corresponds to the
minimum found for given $\gamma$, where the dashed part represents the non-symmetric regime. We see that, differently from what happens for the dependence
on the coupling parameter $g$, the behaviour of the large deviation function against $\gamma$ is
more complex. Globally we can distinguish three regimes: (i)~A weak coupling regime where the large deviation function is symmetric with $s_{\min}$ close to zero; (ii)~A strong coupling regime where a FT is found to hold, with a symmetry point tending to $s_1 = \hbar\omega/k_BT_1$ when the coupling strength $\gamma$ increases; (iii)~An intermediate regime where a FT is not necessarily defined and the symmetric point is rapidly changing with $\gamma$. These three regions are connected to the ones found in Ref.~\cite{Velizhanin2013}, which discusses the heat conduction across a similar system. The scaling behaviour of the mean energy exchange with a given bath $r$ ($\langle K_r\rangle/t=\partial_{s}\theta_r(s)\vert_{s=0}$) for small and large coupling is found in our work to scale respectively as $\gamma$ and $1/\gamma$, matching the results in Ref.~\cite{Velizhanin2013}.

\begin{figure}
\includegraphics[width=0.4\textwidth]{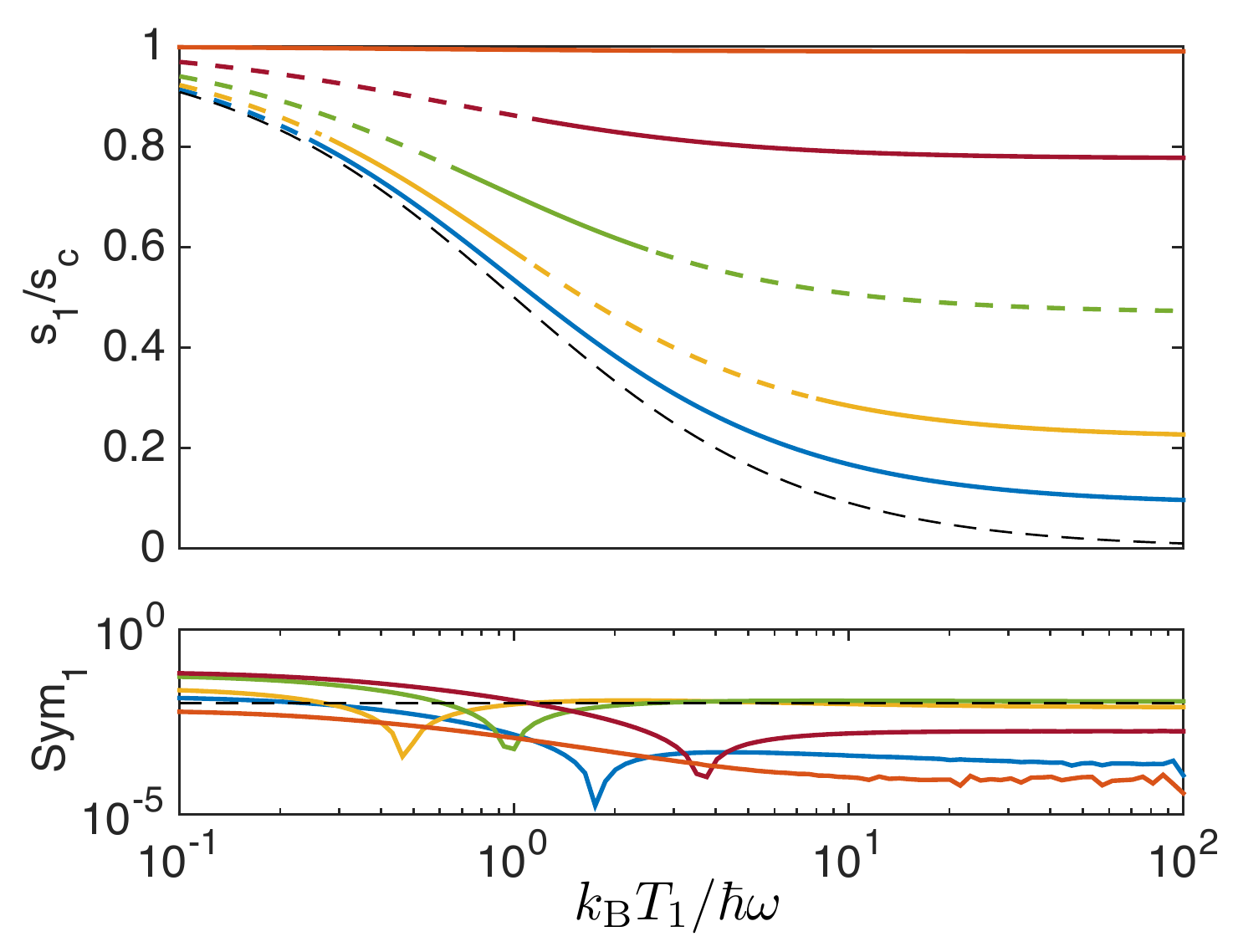}
\caption{Symmetry of the large deviation function $\theta_1(s)$.
In the upper panel the potential FT $s_1=2s_{\min}$ normalised. The thin black
dashed line indicates the result obtained for RW-coupling between oscillators (see section
\ref{sub:32}), while the other curves correspond to different values
of $\gamma$ with from blue to green $\gamma=0.1\omega$, $0.5\omega$, $\omega$, $2\omega$ and $10\omega$. The full line corresponds
to the situation where $s_{1}$ is found to correspond to a FT, while
the dashed part to the situation where it is not, in the sense of the symmetry criterion $\Sym_1$ [\eref{eq:crit}], as represented on the lower panel.\label{fig:thetaxx22}}
\end{figure}

The dependence on the bath temperature is presented in \fref{fig:thetaxx22}, where we focus the attention to potential FTs. The various curves shown correspond to different values of $\gamma$, and we see that we recover the three regimes observed in \fref{fig:thetaxx-1}. For small $\gamma$ (blue curve) we have a defined FT close to the result obtained for the RW-type of coupling (with a finite difference similar to the one previously observed in \fref{fig:thetaxx2}), as expected. The intermediate regime presents less well-defined characteristics (yellow, green and maroon curves).
The impact of the temperature appears to modify importantly the existence of a FT. In general, we can see that for high temperatures the range of existence of a FT tends to be enlarged. For lower temperatures however there is also a defined FT.
Notice that from distinct symmetry criteria, different results may be found, but giving a qualitatively similar picture.
Finally for strong damping we have a defined FT, tending to $s_1\to\hbar\omega/k_BT_1$. In this regime the damping is so strong that the statistics of exchange is only directed by the related bath: The two oscillators appear to be uncoupled.

\section{Conclusion}

We have presented a general framework to fully characterise of excitation-exchange processes between a harmonic network and its environment. The method applies for any network of oscillators with linear and bilinear network interactions, connected to many baths (whether thermal or squeezed) and gives access to the large deviation function attached to a counting process corresponding to the net number of excitations exchanged between the system and a given bath. After giving details of the framework we focused on the possibility, given a large deviation function, to derive local fluctuation theorems related to the exchange with a given bath. After discussing the meaning of these theorems we explored different basic networks: from a single harmonic oscillator to a chain, considering various coupling schemes between oscillators. We found that for the systems considered a local FT can generally be found, especially in the case of RW-like coupling. However position--position coupling can lead, depending on the parameters, to a situation where local fluctuation theorems cannot be defined.

The great versatility of the proposed method applicable to many photonic, phononic and hybrid quantum systems, enable to fully characterise the thermodynamics of exchange taking place with an environment.

\acknowledgments

This work was supported by the UK EPSRC (EP/L005026/1 and EP/J009776/1), the John Templeton Foundation (grant ID 43467), the EU Collaborative Project TherMiQ (Grant Agreement 618074), and the Royal Commission for the Exhibition of 1851. Part of this work was supported by COST Action MP1209 ``Thermodynamics in the quantum regime.'' \\

\appendix*
\section{Phase-space representation of linear and bilinear dynamics of quantum oscillators}\label{appendix}

Here we explicit the derivation of the phase space representation for the different terms of the dynamics as defined in \sref{sub:21}, in terms of the symmetrically-ordered generating function as defined in \eref{eq:generate}.

Considering the definition of the Hamiltonian dynamics involving a single oscillator, as defined in \eref{eq:H}, we have
\begin{widetext}

\begin{equation}
\text{Tr}\left\{ -\tfrac{\imath}{\hbar}[\hat{H}_{i},\hat{\rho}_{s}]e^{\imath\left(\beta_{i}^{*}\hat{a}_{i}^{\dagger}+\beta_{i}\hat{a}_{i}\right)}\right\}  = \Bigl[\imath\hbar\omega_{i}\left(\beta_{i}^{*}\partial_{\beta_{i}^{*}}-\beta_{i}\partial_{\beta_{i}}\right)-2\hbar\omega_{i}d_{i}(t)\left(\beta_{i}^{*}-\beta_{i}\right)  -2\imath\Upsilon_{i}^{*}\beta_{i}\partial_{\beta_{i}^{*}}+2\imath\Upsilon_{i}\beta_{i}^{*}\partial_{\beta_{i}}\Bigr]\chi_{s}\,.\label{eq:H_sp}
\end{equation}
For the dissipative part, as defined in \erefs{eq:lcav-1} and \ref{eq:lcav-2-1}, we have
\begin{eqnarray}
\text{Tr}\left\{ \mathcal{L}_{i}[\hat{\rho}_{s}]e^{\imath\left(\beta_{i}^{*}\hat{a}_{i}^{\dagger}+\beta_{i}\hat{a}_{i}\right)}\right\}  & = & \left[-(\Gamma_{i}-\bar{\Gamma}_{i})\left(\beta_{i}^{*}\partial_{\beta_{i}^{*}}+\beta_{i}\partial_{\beta_{i}}\right)-(\Gamma_{i}+\bar{\Gamma}_{i})\beta_{i}\beta_{i}^{*}\right]\chi_{s}\,,\text{ and}\\
\text{Tr}\left\{ \mathcal{S}_{i}[\hat{\rho}_{s}]e^{\imath\left(\beta_{i}^{*}\hat{a}_{i}^{\dagger}+\beta_{i}\hat{a}_{i}\right)}\right\}  & = & \left(\Lambda_{i}^{*}\beta_{i}^{*}\beta_{i}^{*}+\Lambda_{i}\beta_{i}\beta_{i}\right)\chi_{s}\,.
\end{eqnarray}
Concerning the coupling part between oscillators we have for the different scenarios considered: (i)~Position--position coupling, as defined in \eref{eq:XX}
\begin{equation}
\text{Tr}\left\{ -\tfrac{\imath}{\hbar}[\hat{H}_{ij}^{xx},\hat{\rho}_{s}]e^{\imath\sum_{a=\{i,j\}}\left(\beta_{a}^{*}\hat{a}_{a}^{\dagger}+\beta_{a}\hat{a}_{a}\right)}\right\} =-\imath g_{ij}\left[\left(\beta_{i}-\beta_{i}^{*}\right)\left(\partial_{\beta_{j}}+\partial_{\beta_{j}^{*}}\right)+\left(\beta_{j}-\beta_{j}^{*}\right)\left(\partial_{\beta_{i}}+\partial_{\beta_{i}^{*}}\right)\right]\chi_{s}\label{eq:XX_ps}\,,
\end{equation}
(ii)~RW coupling [\eref{eq:RWA}]
\begin{equation}
\text{Tr}\left\{ -\tfrac{\imath}{\hbar}[\hat{H}_{ij}^{RW},\hat{\rho}_{s}]e^{\imath\sum_{a=\{i,j\}}\left(\beta_{a}^{*}\hat{a}_{a}^{\dagger}+\beta_{a}\hat{a}_{a}\right)}\right\} =-\imath g_{ij}\left(\beta_{i}\partial_{\beta_{j}}-\beta_{i}^{*}\partial_{\beta_{j}^{*}}+\beta_{j}\partial_{\beta_{i}}-\beta_{j}^{*}\partial_{\beta_{i}^{*}}\right)\chi_{s}\label{eq:RWA_ps}\,,
\end{equation}
and (iii)~OPO-like coupling, as defined in \eref{eq:OPO}
\begin{equation}
\text{Tr}\left\{ -\tfrac{\imath}{\hbar}[\hat{H}_{ij}^{OPO},\hat{\rho}_{s}]e^{\imath\sum_{a=\{i,j\}}\left(\beta_{a}^{*}\hat{a}_{a}^{\dagger}+\beta_{a}\hat{a}_{a}\right)}\right\} =-\imath g_{ij}\left(\beta_{i}\partial_{\beta_{j}^{*}}-\beta_{i}^{*}\partial_{\beta_{j}}+\beta_{j}\partial_{\beta_{i}^{*}}-\beta_{j}^{*}\partial_{\beta_{i}}\right)\chi_{s}\label{eq:OPO_ps}\,.
\end{equation}
For what concerns the biased part of the dynamics defined in \eref{eq:ls-2} we have
\begin{equation}
\text{Tr}\left\{ \mathcal{L}_{s}[\hat{\rho}_{s}]e^{\imath\sum_{i=1}^{N}\left(\beta_{i}^{*}\hat{a}_{i}^{\dagger}+\beta_{i}\hat{a}_{i}\right)}\right\}  = -2f_{i+}(s)\left(\partial_{\beta_{i}}\partial_{\beta_{i}^{*}}+\tfrac{1}{4}\beta_{i}\beta_{i}^{*}\right)\chi_{S}(...,\beta_{i},\dots)  -f_{i-}(s)\left(\beta_{i}^{*}\partial_{\beta_{i}^{*}}+\beta_{i}\partial_{\beta_{i}}+1\right)\chi_{S}(...,\beta_{i},\dots)\,,\label{eq:bias_ps}
\end{equation}
\end{widetext}
where $i$ refers here to the bath of reference (from which the counting process $K_i$ is defined).
Combining the above equations we can define the Fokker--Planck equation
ruling the dynamics of the generating function $\chi_{s}$. Decomposing
$\beta_{i}$ in terms of real and imaginary parts, i.e., $\beta_{i}=p_{i}+iq_{i}$,
we can write the Fokker--Planck equation into a matrix form as a function of the vector
$\mathbf{p}^{T}=\left(p_{1},q_{1},...,p_{N},q_{N}\right)$ and corresponding derivative vector
$\partial_{\mathbf{p}}^{T}=\left(\partial_{p_{1}},\partial_{q_{1}},\dots,\partial_{p_{N}},\partial_{q_{N}}\right)$ as presented in \eref{eq:xs}. Notice that the moments $p_{i}$ and $q_{i}$ are respectively
related to the position ($x_{i}$) and momentum ($y_{i}$) of the
$i$th oscillator.

\bibliographystyle{plain}

\end{document}